\documentclass[reprint,prb,showpacs,superscriptaddress]{revtex4-1}
\usepackage{bm}
\usepackage{graphicx}
\usepackage{amsmath,amsfonts,amssymb}
\usepackage{hyperref}
\begin{document}

\title{Section Chern number for a 3D photonic crystal and
the bulk-edge correspondence}
\author{Shuhei Oono}\email{oono@rhodia.ph.tsukuba.ac.jp}
\affiliation{Graduate School of Pure and Applied Science, University of
Tsukuba, Tsukuba 305-8577, Japan}
\author{Toshikaze Kariyado}
\affiliation{Graduate School of Pure and Applied Science, University of
Tsukuba, Tsukuba 305-8577, Japan}
\affiliation{International Center for Materials Nanoarchitectonics, National
Institute for Materials Science, Tsukuba 305-0047, Japan}
\author{Yasuhiro Hatsugai}
\affiliation{Graduate School of Pure and Applied Science, University of
Tsukuba, Tsukuba 305-8577, Japan}

\begin{abstract}
We have characterized the robust propagation modes of electromagnetic waves in helical structures
by the section Chern number that is defined for a two-dimensional (2D) section
 of the three-dimensional (3D) Brillouin zone.
The Weyl point in the photonic bands is
associated with a discontinuous jump of the section Chern number.
A spatially localized
 Gaussian basis set is used to calculate the section
 Chern numbers where we have implemented the divergence-free condition
 on each basis function in 3D.
The validity of the bulk-edge correspondence in
a 3D photonic crystal is discussed in relation to the broken inversion symmetry.
\end{abstract}

\maketitle

\section{Introduction}
Photonic crystals\cite{Yablonovitch1987,Joannopoulos_2008,Sakoda2013}
are systems with a spatially periodic structure of refractive medium.
Electromagnetic (EM) fields in a photonic
crystal are given by the Bloch states
where 
frequency dispersion may have an energy gap as the
electronic band dispersion of solids.
This gives
us an opportunity to simulate microscopic quantum states of electrons by
the macroscopic EM waves governed by the classical Maxwell equations.
Especially, simulating topologically nontrivial phases
in quantum solids has been an extensively studied topic these days. A
prototypical topological phase of matter is an integer quantum Hall
state\cite{Klitzing1980}, where its topological nature is encoded in a topological
invariant, the Chern number, calculated using the Bloch wave
functions\cite{Thouless1982}. A physical consequence of the topological nature is
quantization of the Hall conductance, since it is proportional to the
Chern number. On the other hand, the quantization is also explained in
terms of the chiral edge states\cite{Laughlin1981,Hatsugai1993a}. Generally, there is an intimate
relation between bulk topology and edge states, as known as the
bulk-edge correspondence\cite{Hatsugai1993b}. So, we can access topological properties of
the bulk via edge states, or inversely, we can know about the
edge/surface states by the bulk information alone.

The idea of the bulk-edge correspondence is especially
important in photonic systems
\cite{Raghu_2008,Z.Wang_2008,Z.Wang_2009,Ochiai09,Hafezi11,Fang12,Hafezi13,Khanikaev13,Rechtsman2013,L.Ling2013,LuLing14,L.Ling2015,Wu15},
since there is no ``Hall conductance''
in photonic systems, while edge states are always physical observable. Since
we have the Bloch states and the Brillouin zone in
periodic photonic systems, the
Chern number is well-defined as in electronic systems. Then, comparing
the bulk Chern number and the edge states is a natural
strategy to attack topological issues in photonic systems. In fact,
quantum Hall state analogs in 2D photonic crystals have been discussed
focusing on the edge states. Note that in such studies, the time
reversal symmetry (TRS) breaking
is mandatory to realize a quantum Hall analog.

In this paper, a Gaussian type localized basis set based
method is introduced for numerical evaluation of the Bloch states of the
EM waves. With the localized basis set, momentum, which defines the
Brillouin zone, is treated as a twisted boundary condition
$e^{i\phi_\alpha}$, $\alpha=x,y$, and then, the basis functions
naturally and strictly becomes periodic in momentum $\phi_\alpha$. Since we
will explain later, this feature is advantageous in the Chern number
calculation, and this is the reason for our choice of the localized
basis set.
The Gaussian expansion is not the only choice for the
localized basis set, i.e., for instance, the finite element method 
\cite{Jin2014,Axmann1999,Dobson1999,Dobson2000,Burger2005}
gives an
alternative. However, the Gaussian basis element is convenient 
when derivatives of the basis functions are required, obviously due to
its Gaussian nature. Such a situation really arises in 3D crystals where
the decomposition of the EM field into TE or TM modes is impossible. We
demonstrate the merit of the Gaussian expansion through the
calculation of the ``section'' Chern numbre
\cite{Avron1983,Halperlin1987,Kohmoto1992,Hatsugai04} of the 3D photonic
crystals.

As we explained, the introduced method is applied to 3D
photonic crystals. First of all, note that it is possible to define and
use the Chern number to characterize the given system even in 3D cases
as follows.
That is, 
let us fix one of the three
components of the momentum, say $k_3$,
and define the Chern number $C (k_3)$  using 
$(k_1,k_2)$ as a periodic parameter space.
Since this parameter space is a 2D section of the 3D Brillouin zone,
we call $C (k_3)$
the section
Chern number. This section Chern number is well defined only
when the band gap is always finite on the constant $k_3$ plane (section).
Different from the usual Chern number in 2D
systems, the section Chern number can be finite even in systems having
TRS. This feature enables us to realize a topologically nontrivial state
in photonic crystals without magneto-optical media. Specifically,
$C (k_3)$ can take a nonzero value when the TRS \textit{or} the spatial
inversion symmetry (SIS) is broken. 

Since the section Chern number $C (k_i)$
is a topological invariant, it
can change only
when the band gap vanishes at a certain point on the constant $k_i$
plane. Generally, this gap
closing point is a Weyl
point in the band structure, having a linear dispersion around
it. Intriguing
topological properties are expected in the system with the Weyl
points\cite{Murakami07,X.Wan2011,Burkov2011,Burkov11b}. Also, it has been already shown that the Weyl points
emerge in the  double gyroid photonic crystal when TRS or SIS is broken
\cite{L.Ling2013,L.Ling2015}.

In this paper, we consider a photonic crystal of a simple structure
with Weyl points to discuss the topological nature. In this paper, we
limit ourselves to the cases with TRS but without SIS. 
The considered system is 
anisotropic and the Weyl points
are clearly resolved only
in a constant $k_z$ plane in the Brillouin zone.
We also discuss the edge states associated with the 
nonzero section Chern number, which can be realized by
making a wave packet.

This paper is organized as follows. In Sec.~II, we introduce the
Gaussian expansion method for the photonic crystals, and explain the importance of the
localized basis set for the calculation of the Chern number. Section~III is devoted to demonstrate the validity
and the usefulness of the introduced method by applying the method to
simple models. In Sec.~IV, a specific photonic crystal having the Weyl
points in its photonic band structure is proposed, and the section Chern
number for that model is evaluated. The relation between the section
Chern number and the edge modes is also discussed. Summary and discussion are
given in Sec.~V.

\section{Methods}

\begin{figure}[tbp]
\begin{center}
 \includegraphics[width=\columnwidth]{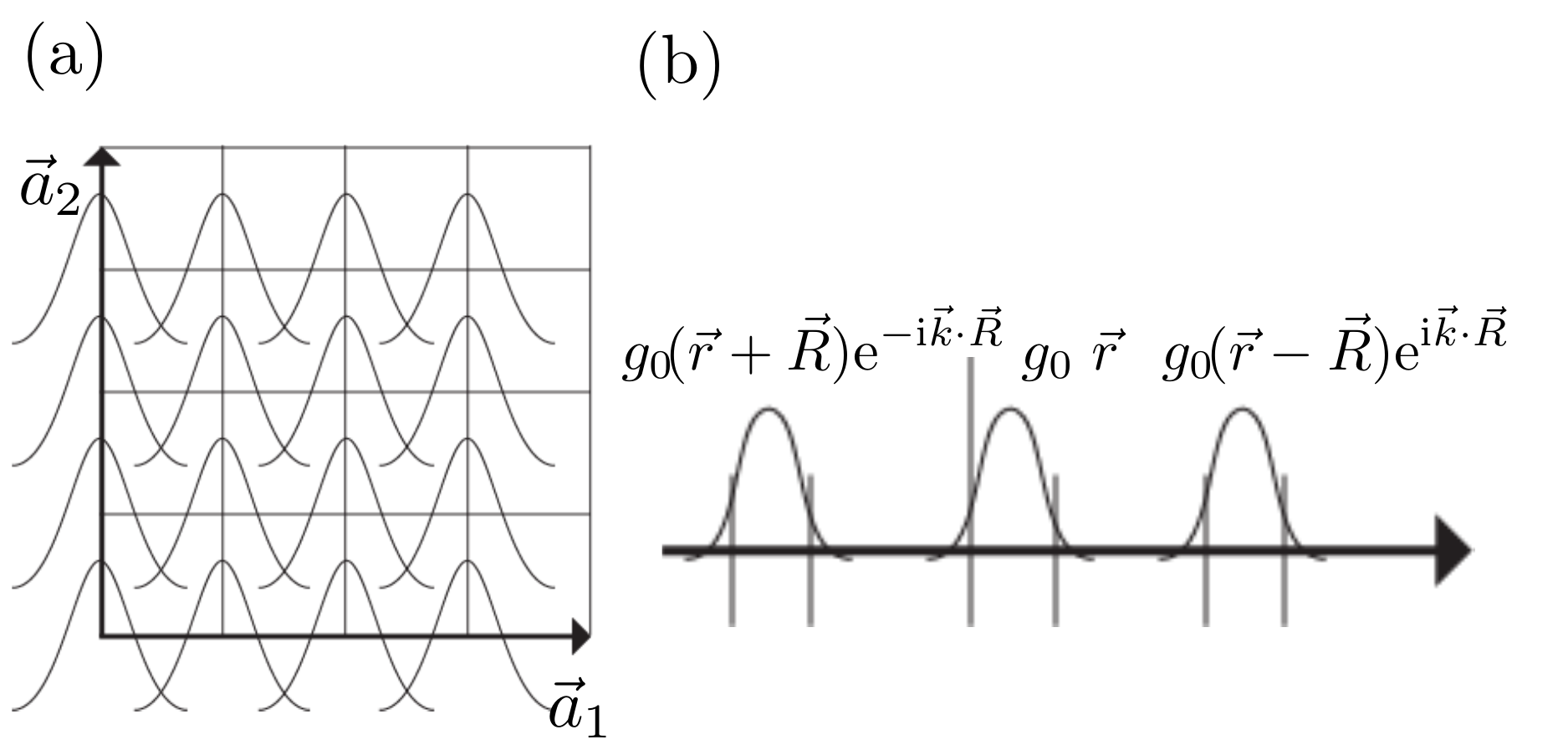}
 \caption{(a) Gaussian (spatially localized) basis elements are put on the grid that divides unit cell
 uniformly in the directions of $\boldsymbol{a}_1$ and
 $\boldsymbol{a}_2$. (b) Each basis element is set on each unit cell with the
 factor {$\mathrm{e}^{\mathrm{i} \boldsymbol{k} \cdot \boldsymbol{R}}$}.} \label{gaussian-picture}
\end{center} 
\end{figure}

The Maxwell equations for the normal modes introduced as $\boldsymbol{E}
\negthinspace \left( \boldsymbol{r}, t \right) =
{\mathrm{e}^{-\mathrm{i} \omega t}}  \boldsymbol{E}
\negthinspace \left( \boldsymbol{r} \right)$ and $\boldsymbol{H}
\negthinspace \left( \boldsymbol{r}, t \right) =
{\mathrm{e}^{-\mathrm{i}\omega t}} \boldsymbol{H} \negthinspace \left( \boldsymbol{r} \right)$ are written as 
\begin{equation}
 \begin{aligned}
  \nabla \times \boldsymbol{E} \negthinspace \left( \boldsymbol{r}
  \right) &= {\mathrm{i} \omega}
 \boldsymbol{B}(\bm{r}), & 
  \nabla \times \boldsymbol{H} \negthinspace \left( \boldsymbol{r}
  \right) &= 
{-\mathrm{i} \omega }
\boldsymbol{D}(\bm{r}), \\
  \nabla \cdot \boldsymbol{D} \negthinspace \left( \boldsymbol{r} \right)
  &= 0, & \nabla \cdot \boldsymbol{B} \negthinspace \left( \boldsymbol{r} \right)
  &= 0. \label{Maxwell-eq}
 \end{aligned}
\end{equation} 
Here, we assume that the permittivity and the permeability are linear and
lossless, i.e., we assume
\begin{equation}
 \boldsymbol{D} \negthinspace \left( \boldsymbol{r} \right) =
  \varepsilon_0 \hat{\varepsilon} \negthinspace \left( \boldsymbol{r} \right) \boldsymbol{E}
  \negthinspace \left( \boldsymbol{r} \right), 
\thickspace 
 \boldsymbol{B} \negthinspace \left( \boldsymbol{r} \right) =
  \mu_0 \hat{\mu} \negthinspace \left( \boldsymbol{r} \right) \boldsymbol{E}
  \negthinspace \left( \boldsymbol{r} \right),
\end{equation}
where $\hat{\varepsilon} \negthinspace \left( \boldsymbol{r} \right)$
and $\hat{\mu} \negthinspace \left( \boldsymbol{r} \right)$ are $3 \times
3$ tensors satisfying the Hermitian conditions
$\hat{\varepsilon}^\dagger = \hat{\varepsilon}$ and $\hat{\mu}^\dagger =
\hat{\mu} $. With this assumption, the Maxwell equations are rewritten
into Hermitian eigenequation forms as
\begin{align}
\nabla \times \hat{\varepsilon}^{-1}\negthinspace \left( \boldsymbol{r} \right) \nabla \times \boldsymbol{H}
 \negthinspace \left( \boldsymbol{r} \right) &= \left( \frac{\omega}{c}
					       \right)^2 
\hat{\mu} \negthinspace \left( \boldsymbol{r} \right) \boldsymbol{H}
\negthinspace \left( \boldsymbol{r} \right), \label{eigenequation-H}\\
\nabla \times \hat{\mu}^{-1}(\bm{r}) \nabla \times \boldsymbol{E}
 \negthinspace \left( \boldsymbol{r} \right) &= \left( \frac{\omega}{c}
					       \right)^2 
\hat{\varepsilon} \negthinspace \left( \boldsymbol{r} \right) \boldsymbol{E}
\negthinspace \left( \boldsymbol{r} \right). \label{eigenequation-E}
\end{align}
We only have to solve the  either of these eigenequations with respect to
$\boldsymbol{E}$ or $\boldsymbol{H}$.
In a spatially periodic system, all the eigenstates are
represented as Bloch states. 

Roughly speaking, there are two ways to expand a
Bloch state. The first way is to use a basis set localized in wave space
like plane waves
\cite{Ho90,Plihal1991,Sozuer1993,Meade93,Johnson01}, and the second way is to use a basis set localized in
real space.
For the plane wave type basis sets
, 
a Bloch state is usually represented as
\begin{equation}
 \psi_{k} \negthinspace \left( \boldsymbol{r} \right) = \sum_G f_G
{\mathrm{e}^{\mathrm{i} \left(\boldsymbol{G} 
+ \boldsymbol{k} \right)\cdot \boldsymbol{r} }}
, \label{plane-wave}
\end{equation}
with $\boldsymbol{G}$ denoting reciprocal lattice vectors. In this form,
each basis element is not periodic for a parameter change $\boldsymbol{k}
\rightarrow \boldsymbol{k} + \boldsymbol{G} $, since
{$\mathrm{e}^{\mathrm{i} \boldsymbol{k} \cdot \boldsymbol{r} }$}
 and
{$\mathrm{e}^{\mathrm{i} \left( \boldsymbol{G} + \boldsymbol{k}\right)
\cdot \boldsymbol{r} }$ }
are not identical.

On the other hand, for spatially localized basis sets, the matrix 
eigenequation can be naturally made into periodic in parameter
$k$. A Bloch state is represented by localized basis set as
\begin{equation}
 \psi_k \negthinspace \left( \boldsymbol{r} \right) = \frac{1}{\sqrt{N}}
  \sum_m 
{\mathrm{e}^{\mathrm{i} \boldsymbol{k} \cdot \boldsymbol{R}_m }}
  w_m \negthinspace \left(\boldsymbol{r} \right), \label{localize-wavenumber}
\end{equation}
with $\boldsymbol{R}_m $ and $N$ representing a lattice vector in real
space and the number of unit cells of the system under consideration
respectively, and $w_m \negthinspace \left(
\boldsymbol{r} \right) = w_0 \negthinspace \left( \boldsymbol{r} -
\boldsymbol{R}_m \right)$. In this form, a parameter $k$ only appears as
{$\boldsymbol{e}^{\mathrm{i} \boldsymbol{k} \cdot
\boldsymbol{R}_m }$} 
and is introduced in the eigenequation as a boundary condition.
This periodicity is crucial in the numerical evaluation of the Chern
number\cite{Fukui2005}. 

Representative localized functions are the maximally localized Wannier
functions\cite{Marzari1997,Whittaker2003,Busch2003,Marzari2014}, however,
it is known that the so-called \textit{composite bands}, which compose
the maximally localized
Wannier functions, must be distinguished from the bands below and above by
gaps with vanishing Chern number
\cite{Panati2007,Brouder2007}.
However, photonic bands are generally tangling in
higher frequencies and it is not common for photonic band structures
to have an isolated set of bands enclosed by gaps.

Then, we use a Gaussian basis set as a typical
localized basis set in the following. Explicitly, the Gaussian
 basis element is written as 
\begin{equation}
 \boldsymbol{g}_{i,k}^{\lambda} \negthinspace \left(
			\boldsymbol{r} \right) =
\frac{1}{\sqrt{N}} \sum_{m,\lambda} 
{\mathrm{e}^{\mathrm{i} \boldsymbol{k} \cdot \boldsymbol{R}_m}}
\boldsymbol{f}^{\lambda} \negthinspace \left( \boldsymbol{r}_{im} \right)
\mathrm{e}^{-  r_{im}^2
/ \alpha^2}, \label{gaussian-wavenumber}
\end{equation}
with
\begin{equation}
 \bm{f}^{ \lambda} (\bm{r}_{im})=
 \begin{pmatrix}
  f_x^{ \lambda} \negthinspace \left(x_{im},y_{im},z_{im} \right)\\
  f_y^{ \lambda} \negthinspace \left(x_{im},y_{im},z_{im} \right)\\
  f_z^{ \lambda} \negthinspace \left(x_{im},y_{im},z_{im} \right)
 \end{pmatrix}.
\end{equation}
Here, suffices $i$ and $ \lambda$ specify the grid position within the unit cell
and the polarization degrees of freedom respectively, and $\boldsymbol{r}_{im}
= \boldsymbol{r} - \left( \boldsymbol{R}_m + \boldsymbol{r}_i
\right)$. The $f_i^{ \lambda}\negthinspace \left( x, y, z \right)$
is a polynomial of $\left(x, y, z\right)$. 
Using this basis set for the eigenfunction expansion in Eq.~\eqref{eigenequation-H}, we obtain a
matrix eigenequation 
\begin{equation}
\sum_{j,\lambda'}
 \Theta_{i\lambda,j
 \lambda'} \negthinspace \left( \boldsymbol{k} \right) c_{j}^{\lambda'} =
 \left( \frac{\omega}{c} \right)^2
 \sum_{j,\lambda'}
 O_{i\lambda,j\lambda'} \negthinspace \left(
  \boldsymbol{k} \right) c_{j}^{\lambda'}. \label{eigenequation-matrix}
\end{equation}
Here, $\Theta_{i\lambda,j\lambda'} \negthinspace \left(\boldsymbol{k}\right)$ is the
matrix element of the operator in the left hand side of 
Eq.~\eqref{eigenequation-H}, i.e., $\hat{\Theta} = \nabla \times
\hat{\varepsilon}^{-1} \negthinspace \left( \boldsymbol{r} \right)
\nabla \times$, which is explicitly written as
\begin{eqnarray}
  \Theta_{i\lambda,j\lambda'} \negthinspace \left( \boldsymbol{k} \right) &&=
\left< \boldsymbol{g}_{ik}^{ \lambda} \right| \nabla \times \hat{\varepsilon}^{-1}
  \negthinspace \left( \boldsymbol{r} \right) \nabla \times \left|
		 \boldsymbol{g}_{jk}^{\lambda'} \right> \notag \\
&&= \sum_{m} 
{\mathrm{e}^{ -\mathrm{i} \boldsymbol{k} \cdot  \boldsymbol{R}_m }}
\int \mathrm{d}^3 r \left( \nabla \times \boldsymbol{f}^{\lambda} \negthinspace \left( \boldsymbol{r}_{im} \right)  \mathrm{e}^{- r_{im}^2/\alpha^2 }\right)^\ast \notag \\
&& \qquad \qquad
\cdot \hat{\varepsilon}^{-1}
  \negthinspace \left( \boldsymbol{r} \right)
\left( \nabla \times
  \boldsymbol{f}^{\lambda} \negthinspace \left( \boldsymbol{r}_{j0} \right)
  \mathrm{e}^{- r_{j0}^2 / \alpha^2 }\right). \label{matrix-element-theta}
\end{eqnarray}
Here we used the Hermiticity of the rotation,
$
\int \mathrm{d}^3 r \thickspace \boldsymbol{g}^\ast \negthinspace \left(
							       \boldsymbol{r}\right) 
\cdot \left\{ \nabla \times \boldsymbol{f} \negthinspace \left( \boldsymbol{r}\right) \right\}
=
 \int \mathrm{d}^3 r \thickspace \boldsymbol{f} \negthinspace \left(
							       \boldsymbol{r}\right) 
\cdot \left\{ \nabla \times \boldsymbol{g} \negthinspace \left(
							  \boldsymbol{r}\right)
      \right\}^\ast
$ under the boundary condition, 
$\displaystyle \lim_{r \rightarrow \infty }
\boldsymbol{g}_{i,k}^{ \lambda} \negthinspace \left(
			\boldsymbol{r} \right) = 0$. 
On the other hand,
$O_{i\lambda,j\lambda'} \negthinspace \left( \boldsymbol{k} \right)$ is the
overlap, which is obtained as,
\begin{eqnarray}
  O_{i\lambda,j\lambda'} \negthinspace \left( \boldsymbol{k} \right) &&=
\left< \boldsymbol{g}_{ik}^{\lambda} \right| \hat{\mu} \negthinspace \left( \boldsymbol{r}
  \right) \left| \boldsymbol{g}_{jk}^{\lambda'} \right>  \notag \\
&&= \sum_{m} 
{\mathrm{e}^{ -\mathrm{i} \boldsymbol{k} \cdot  \boldsymbol{R}_m } }
\int \mathrm{d}^3 r \left( \boldsymbol{f}^{\lambda} \negthinspace \left( \boldsymbol{r}_{im} \right)
  \mathrm{e}^{-
  r_{im}^2 / \alpha^2 } \right)^\ast \notag \\
&& \qquad \qquad
\cdot \hat{\mu}
 \negthinspace \left( \boldsymbol{r} \right)
 \left( \boldsymbol{f}^{\lambda'} \negthinspace \left( \boldsymbol{r}_{j0} \right)
  \mathrm{e}^{-
  r_{j0}^2 / \alpha^2} \right). \label{matrix-element-overlap}
\end{eqnarray}
The integration is performed over the infinite spatial region.
In many cases, a photonic crystal is constituted by repetition of
structures having different permittivity or permeability from the
uniform media. Then, it is convenient to separate the inverse of
permittivity $\hat{\varepsilon}^{-1} \negthinspace \left( \boldsymbol{r}
\right)$ as 
\begin{equation}
\hat{\varepsilon}^{-1} \negthinspace \left(
\boldsymbol{r} \right) = \hat{\varepsilon}_c^{-1} + \left( \hat{\varepsilon}^{-1} \negthinspace \left(
\boldsymbol{r} \right) - \hat{\varepsilon}_c^{-1} \right) \label{permittivity-separate}.
\end{equation}
The first term of the right hand side is constant corresponding to the
uniform background media. The second term is
nonzero only in the region where the structures exist. For the constant term, the integrals of
the matrix elements are the standard Gaussian integral, which can be analytically
evaluated.
For the second term, the numerical integration is easy since the
integrand is finite only in the limited region.
The same argument also applies to the inverse of the permeability $\hat{\mu}^{-1}
\negthinspace \left(\boldsymbol{r} \right)$. For a uniform system all the matrix elements
are given without numerical integration.

\subsection{Gaussian basis set (2D case)}
First, we consider the 2D case. Here, 2D means that $\hat{\varepsilon}
\negthinspace \left( \boldsymbol{r} \right)$ and $\hat{\mu}
\negthinspace \left( \boldsymbol{r} \right)$ have no dependence on $z$.
In such a system, a mirror plane perpendicular to the $z$ axis exists and the
solutions of Eqs.~\eqref{eigenequation-H} and
\eqref{eigenequation-E}
 are separated into TE modes ($H_z$ polarization) or TM modes ($E_z$
 polarization). As a result, we only have to consider the
$z$ component of either field $\boldsymbol{H}$ or $\boldsymbol{E}$, which indicates that the
 equation to be solved becomes a scalar equation.

For the scalar eigenequation, we use  a basis element 
\begin{equation}
 g_{ik} \negthinspace \left( \boldsymbol{r} \right) = \frac{1}{\sqrt{N}}
  \sum_m 
{\mathrm{e}^{\mathrm{i} \boldsymbol{k} \cdot \boldsymbol{R}_m }}
  \mathrm{e}^{-r_{im}^2 / \alpha^2 }. \label{2d-gaussian-base}
\end{equation}
Each basis function is cylindrically
symmetric and there are finite overlaps between the basis functions.
Then, these  basis elements are
put on  the grid dividing the unit
cell regularly (Fig.\ref{gaussian-picture}).
Later, we will see that 
the simple configuration of the basis elements gives
sufficient accuracy for our purpose, and there arises no need for
further attention on the distribution of the grids, which is often
important in the standard finite element methods.

For this basis set, the integral of the overlap $O_{ij}
\negthinspace \left( \boldsymbol{k} \right)$ for the constant term
(when $\thinspace \hat{\mu}_c = \hat{1}$)
is obtained as
\begin{equation}
O_{ij}^\text{empty}
\negthinspace \left( \boldsymbol{k} \right)
= 
\sum_{m} \frac{\pi \alpha^2 }{2} 
{\mathrm{e}^{-\mathrm{i} \boldsymbol{k} \cdot \boldsymbol{R}_m}}
\mathrm{e}^{- \frac{{r_m'}^2 }{2 \alpha^2} }, 
\label{2d-overlap-value}
\end{equation}
where  $\boldsymbol{r}_m' = \boldsymbol{r}_i + \boldsymbol{R}_m
- \boldsymbol{r}_j$
.
On the other hand, the integral of the matrix
element $\Theta_{ij} \negthinspace \left( \boldsymbol{k} \right)$ for
the constant term (when $\hat{\varepsilon}_c^{-1} = \hat{1} $) is
\begin{equation}
\Theta_{ij}^\text{empty} \negthinspace \left( \boldsymbol{k} \right)
= 
\sum_{m} \frac{\pi \alpha^2}{2}
{\mathrm{e}^{-\mathrm{i} \boldsymbol{k} \cdot \boldsymbol{R}_m}}
\mathrm{e}^{-\frac{{r_m'}^2}{2 \alpha^2} } \frac{1}{\alpha^2}
\left( 2 - \frac{{r_m'}^2}{\alpha^2} \right).\label{2d-matrix-value}
\end{equation}

For the empty lattice, all the matrix elements of
Eq.~\eqref{eigenequation-matrix} are given by using
Eqs. \eqref{2d-overlap-value} and \eqref{2d-matrix-value}. When some
structures are introduced, the additional work to obtain matrix elements
is the numerical integration within the structures.

\subsection{Gaussian basis set (3D case)}
 For a 3D system, it is generally impossible to decompose the 
 EM fields into TE or TM modes, and we have to
 handle all three components of the vector eigenequation. 
A naive thought suggests us to use the scalar
 Gaussian functions [Eq.~\eqref{2d-gaussian-base}] for each component of
 the vector as
\begin{equation}
 \boldsymbol{\psi}_k \negthinspace \left( \boldsymbol{r} \right)
= \sum_{ \lambda = 1}^3
\boldsymbol{e}_{ \lambda}  \sum_i c_{ik}^{\lambda} g_{ik} \negthinspace
\left( \boldsymbol{r} \right) . \label{extensionAtFirst}
\end{equation}
Here, $\boldsymbol{e}_{ \lambda} $ represents unit vectors in three orthogonal
directions. This naively introduced basis set, however,
has a
deficiency in the following sense. Firstly, note that the divergence of the 
Eq.~\eqref{eigenequation-H} vanishes
because of $\nabla \cdot \left( \nabla \times \boldsymbol{f}
\negthinspace \left( \boldsymbol{r} \right) \right) = 0$ for any
vector field $\boldsymbol{f} \negthinspace \left( \boldsymbol{r}
\right)$. Thus, when the permeability is constant and 
isotropic, the solution of
Eq.~\eqref{eigenequation-H} satisfies the constraint,
\begin{equation}
 \nabla \cdot \boldsymbol{H} \negthinspace \left( \boldsymbol{r} \right)
  = 0  \label{constraint}.
\end{equation}
On the other hand, the
divergence of Eq.~\eqref{extensionAtFirst} is
\begin{equation}
 \nabla \cdot \sum_{i, \lambda}
  \boldsymbol{e}_{ \lambda}  c_{ik}^{\lambda} g_{ik} \negthinspace
\left( \boldsymbol{r} \right) = \sum_{i,\lambda} 
\left( - 2 \alpha x_{\lambda} \right)  c_{ik}^{\lambda} g_{ik} \negthinspace \left(
 \boldsymbol{r} \right),
\end{equation}
where the right-hand side does not vanish except in the trivial case
with  $c_{ik}^{ \lambda} = 0$ for all combinations of $i$ and ${\lambda}$.
Namely, expansion Eq.~\eqref{extensionAtFirst} does not
satisfy constraint Eq.\eqref{constraint}, and as a consequence, the spectrum of the
eigenvalues includes unphysical spurious values. 
For each basis to satisfy constraint Eq.~\eqref{constraint}, we modify each
basis by taking its rotation as
\begin{equation}
\begin{split}
 \boldsymbol{g}_{ik}^{\lambda} \negthinspace
\left( \boldsymbol{r}\right) 
&= \nabla \times \boldsymbol{e}_{\lambda} g_{ik} \negthinspace
  \left( \boldsymbol{r} \right) \\
&= \sum_{ \mu, \nu = 1}^3
 \mathrm{\varepsilon}_{\mu \lambda
 \nu} \left( - 2 \alpha
x_{\mu} \right) g_{ik} \negthinspace \left( \boldsymbol{r} \right) 
\boldsymbol{e}_{\nu} , \label{the-base}
\end{split}
\end{equation}
where $\varepsilon_{\lambda \mu \nu} $ is an anti-symmetric symbol
$\left(\varepsilon_{123} = 1 \right)$.
This is also obtained by choosing three independent
 basis elements of the lowest
order from Eq.~\eqref{gaussian-wavenumber} under the divergence-free condition.
The modified basis element Eq.~\eqref{the-base} has no divergence, and satisfies
the constraint [Eq.~\eqref{constraint}]. We expand the solution
using these basis elements as
\begin{equation}
 \boldsymbol{\psi}_k \negthinspace \left( \boldsymbol{r} \right)
= \sum_{{\lambda} = 1}^3  \sum_i c_{ik}^{\lambda} \boldsymbol{g}_{ik}^{\lambda} \negthinspace
\left( \boldsymbol{r} \right) . \label{expansion-vec}
\end{equation}
Here again, we write down the integral values for this basis set.
The integrals for the constant term of the overlaps
$O_{iz,jz} \left( \boldsymbol{k} \right) $ and
$O_{ix,jz} \left( \boldsymbol{k} \right)$ are 
\begin{eqnarray}
O_{iz,jz}^\text{empty} \negthickspace \left( \boldsymbol{k} \right)
&& 
 = \sum_{m} \sqrt{\left(\frac{\pi \alpha^2}{2} \right)^3} 
{\mathrm{e}^{-\mathrm{i} \boldsymbol{k} \cdot \boldsymbol{R}_m }}
 \mathrm{e}^{-\frac{{r_m'}^2}{2 \alpha^2} }
\notag \\ 
&& 
\qquad \qquad \qquad \times
\frac{1}{\alpha^2}
\left\{ 2 - \frac{1}{\alpha^2} \left( {x_m'}^2 +
 {y_m'}^2  \right) \right\} 
\label{3d-overlap-value1} \quad \\
O_{ix,jz}^\text{empty} \negthickspace \left( \boldsymbol{k} \right)
&&
 = \sum_m \sqrt{\left(\frac{\pi \alpha^2}{2}\right)^3}
{\mathrm{e}^{-\mathrm{i} \boldsymbol{k} \cdot \boldsymbol{R}_m}}
 \mathrm{e}^{-\frac{{r_m'}^2}{2 \alpha^2} }  \frac{1}{\alpha^4} x_m' z_m'
 \label{3d-overlap-value2}
\end{eqnarray}
respectively, while the integrals for the matrix elements
$\Theta_{iz,jz} \left(\boldsymbol{k}\right)$ and $\Theta_{ix,jz} \left(\boldsymbol{k}\right)$ are
\begin{eqnarray}
&&
\Theta_{iz,jz}^\text{empty} \negthickspace \left(\boldsymbol{k}\right)
= 
\sum_m
\sqrt{\left(\frac{\pi \alpha^2}{2}\right)^3}
{\mathrm{e}^{-\mathrm{i} \boldsymbol{k} \cdot \boldsymbol{R}_m }}
\mathrm{e}^{-\frac{{r_m'}^2}{2 \alpha^2}}
\notag \\
&&
\qquad 
\times
\frac{1}{\alpha^4}
 \Big[ \frac{1}{\alpha^4} \left( {x_m'}^2 + {y_m'}^2 \right)^2
- \frac{9}{\alpha^2} \left( {x_m'}^2 + {y_m'}^2 \right)\notag \\
&& 
\qquad 
\qquad 
+ \frac{{z_m'}^2}{\alpha^2}
\Big\{ \frac{1}{\alpha^2} \left( {x_m'}^2 + {y_m'}^2 \right) -2 \Big\} + 10\Big]
\label{3d-matrix-value1}  \\
&&
\Theta_{ix,jz}^\text{empty} \negthickspace \left(\boldsymbol{k}\right) 
= \sum_m \sqrt{\left(\frac{\pi \alpha^2}{2}\right)^3}
{\mathrm{e}^{-\mathrm{i} \boldsymbol{k} \cdot
\boldsymbol{R}_m } }
\mathrm{e}^{-\frac{{r_m'}^2 }{2 \alpha^2} }
\notag \\
&&
\qquad 
\qquad 
\qquad 
\qquad 
\times
\frac{{x_m'} {z_m'}}{\alpha^6}
\left( 7 - \frac{{r_m'}^2}{\alpha^2} \right)
 \label{3d-matrix-value2} 
\end{eqnarray}
respectively.
The integrals of the other components are obtained by 
permutations of the indices $x$, $y$, and $z$. In addition,
the integrals are invariant against the simultaneous exchange of the
position of the localization center and the direction of the
polarization $\left( i \lambda'
\leftrightarrow j \lambda \right)$.
 Thus all matrix components for 
 an isotropic bulk dielectric are given by
Eqs.~\eqref{3d-overlap-value2}-\eqref{3d-matrix-value2}.
The grid points for these  basis elements can be chosen to divide the unit cell of the Bravais lattice
with a regular spacing like Fig.~\ref{gaussian-picture}. 

Here, we make some comments on the case of spatially varying
$\hat{\mu}(\bm{r})$.
In such a case, we have
$\nabla\cdot\bm{H}(\bm{r})\neq{0}$, but the expansion
 (Eq.~\eqref{extensionAtFirst}) still produces a spurious spectrum.
However, the basis set (Eq.~\eqref{the-base}) is also insufficient because it
has no longitudinal component of $\boldsymbol{H} \negthinspace \left(
\boldsymbol{r} \right)$, which should arise in this case. 

\begin{figure*}[htbp]
\begin{center}
 \includegraphics[width=2\columnwidth]{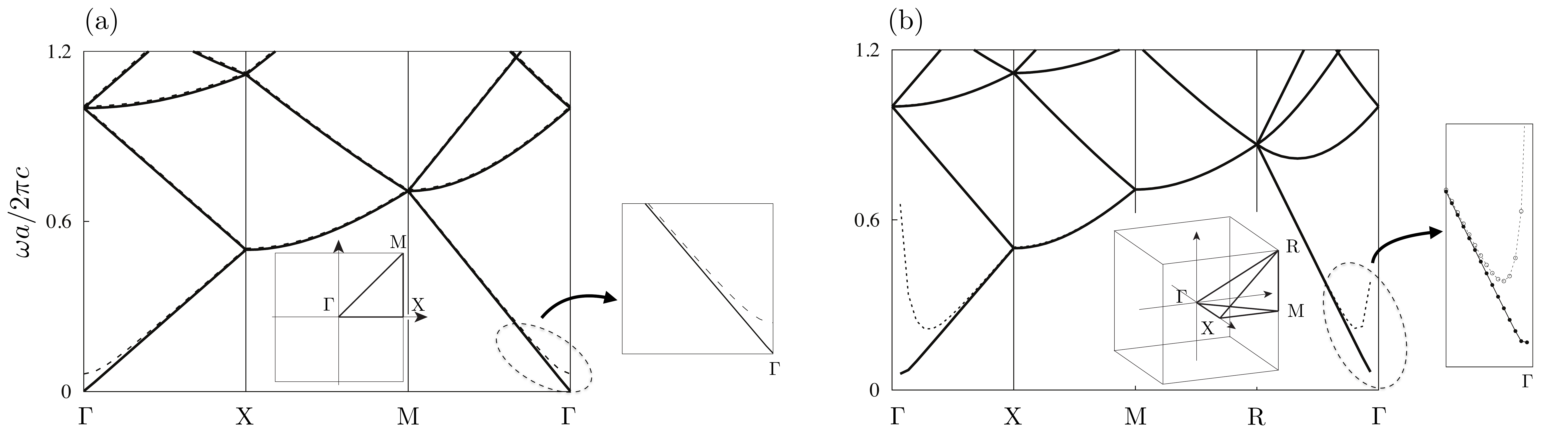}
 \caption{Band structures of empty lattices (homogeneous systems
 $\hat{\varepsilon} = \hat{1}$ and $\hat{\mu} = \hat{1}$). (a)
 Dispersion of 2D square empty lattice calculated by using $64$ ($8
 \times 8$) scalar Gaussian basis elements. The 
 basis elements are located on grid points
 that regularly the divide square lattice. Results of different settings are
 compared between $\alpha = d$ (solid lines) and $\alpha = 0.83 d$ (broken lines). For the basis set with smaller overlap (broken
 line),the lowest eigenvalue at $\Gamma$ point does not converge to
 $0$. (b) Dispersion of 3D simple cubic empty lattice
 calculated by using 1,536  basis elements with
 $\alpha = 1.07  d$ (solid lines) and
 $\alpha = d$ (broken line). The fictitious
 divergence of the lowest eigenvalue near  $\Gamma$ suppressed by
 taking larger overlap (solid line). The lowest eigenstate at $\Gamma$
 point cannot be restored by the vector Gaussian basis
 set.}\label{empty-lattice} 
\end{center} 
\end{figure*}

\subsection{Chern number calculation}
The most sophisticated method for the calculation of the Chern
number seems to be the one devised by Fukui, Hatsugai and
Suzuki\cite{Fukui2005}. Here, let us review the method briefly.
In this method,
the Chern number is derived on a
discretized mesh of
the Brillouin zone.
An $U \negthinspace \left( 1 \right)$ link variable is
defined as
\begin{equation}
%
 U_\mu \negthinspace \left( \boldsymbol{k}_l \right) \equiv 
\frac{
\langle n \negthinspace \left( \boldsymbol{k}_l \right)
| n \negthinspace \left( \boldsymbol{k}_l + \boldsymbol{\mathrm{e}}_\mu \right) \rangle} 
{\left| \langle n \negthinspace \left( \boldsymbol{k}_l \right) 
| n \negthinspace \left( \boldsymbol{k}_l +
			\boldsymbol{\mathrm{e}}_\mu \right) \rangle \right|}
 \label{link-variable},
\end{equation}
where $\left| n \negthinspace \left( \boldsymbol{k}_l \right) \right>$
is the $n$th eigenvector obtained by diagonalizing an eigenequation,
$\boldsymbol{k}_l $ represents a lattice point in the discretized,
Brillouin zone and $\boldsymbol{\mathrm{e}}_\mu$ represents one lattice displacement
in the direction $\mu\left( = 1, 2\right)$. The $\boldsymbol{k}_l$ is
invariant under the displacement of periodic length $\boldsymbol{k}_l +
N_\mu \boldsymbol{\mathrm{e}}_\mu = \boldsymbol{k}_l$. From the link variable, a
lattice field strength is taken as
\begin{eqnarray}
\tilde{F}_{12} \negthinspace \left( \boldsymbol{k}_l \right) &&\equiv \ln 
U_1 \negthickspace \left( \boldsymbol{k}_l  \right)
U_2 \negthickspace \left( \boldsymbol{k}_l \negthickspace +
 \negthickspace \boldsymbol{\mathrm{e}}_1
 \right)
U_1^{\scriptscriptstyle -1} \negthickspace \left( \boldsymbol{k}_l \negthickspace +
 \negthickspace \boldsymbol{\mathrm{e}}_2
 \right)
U_2^{\scriptscriptstyle -1} \negthickspace \left( \boldsymbol{k}_l  \right)
 \notag \\ 
&&- \pi < \frac{1}{\mathrm{i}} \tilde{F}_{12} \negthinspace \left( \boldsymbol{k}_l
 \right) \le \pi. \label{field-strength} 
\end{eqnarray}
Here the field strength is defined as the principal branch of the
logarithm.
From this field strength,
the Chern number is computed as 
\begin{equation}
 \tilde{c} = \frac{1}{2 \pi \mathrm{i} } \sum_l \tilde{F}_{12} \negthinspace
  \left(\boldsymbol{k}_l \right),
\end{equation}
where the sum is taken over all the lattice points in Brillouin
zone. It is straightforward to extend this method to multiband cases.

Equation~\eqref{field-strength} naturally reflects
gauge invariance. The result is always given as an integer
because of the periodicity of the parameters ($| n\negthinspace
\left(\boldsymbol{k}_l + N_\mu \boldsymbol{\mathrm{e}}_\mu \right)
\rangle = | n\negthinspace
\left(\boldsymbol{k}_l \right)
\rangle $) and it can be seen by fixing
the gauge over the entire lattice points. A gauge
potential is defined as
\begin{equation}
 \widetilde{A}_\mu \negthinspace \left(\boldsymbol{k}_l\right) = \ln
  U_\mu \negthinspace \left( \boldsymbol{k}_l\right), \quad - \pi <
  \frac{1}{\mathrm{i}} \widetilde{A}_\mu \negthinspace \left( \boldsymbol{k}_l
						   \right)
 \le \pi,
\end{equation}
which is periodic on the lattice $\tilde{A}_\mu \negthinspace \left(
\boldsymbol{k}_l + N_\mu \boldsymbol{\mathrm{e}}_\mu \right) = \tilde{A}_\mu \negthinspace \left(
\boldsymbol{k}_l \right)$. The field strength (Eq.~\eqref{field-strength}) is
related to this gauge potential by
\begin{equation}
 \tilde{F}_{12} \negthinspace \left(\boldsymbol{k}_l \right) =
  \triangle_1 \widetilde{A}_2 - \triangle_2
  \widetilde{A}_1 + 2 \pi \mathrm{i} n_{12} \negthinspace
  \left( \boldsymbol{k}_l \right), \label{field-strength-gauge-potential}
\end{equation}
where $\triangle_\mu$ represents a forward difference operation
$\triangle_\mu \widetilde{\boldsymbol{A}} \negthinspace \left(
\boldsymbol{k}_l \right) =  \widetilde{\boldsymbol{A}} \negthinspace \left(
\boldsymbol{k}_l + \boldsymbol{e}_\mu \right) -
\widetilde{\boldsymbol{A}} \negthinspace \left(
\boldsymbol{k}_l \right)
$
and $n_{12} \negthinspace \left( \boldsymbol{k}_l \right)$ is an
integer-valued field that makes $ - \pi < \tilde{F}_{12}/\mathrm{i}
\leq
\pi$. Taking the sum of Eq.~\eqref{field-strength-gauge-potential}, the
first and second terms of the right hand side cancel between adjacent
links. 
Then the total sum of those terms vanishes because of the periodicity, and
the result is given by
\begin{equation}
 \tilde{c} = \sum_l n_{12} \negthinspace \left( \boldsymbol{k}_l
					   \right). 
\end{equation}

For the Gaussian(spatially localized) basis sets, $\left| n \negthinspace \left(\boldsymbol{k}
\right) \right>$ is given by the eigenvector of the matrix eigenequation
(Eq.~\eqref{eigenequation-matrix}). Since Eq.~\eqref{eigenequation-matrix} is
invariant under the displacement $\boldsymbol{k} \rightarrow \boldsymbol{k} +
\boldsymbol{G}$, it is guaranteed that the
computed Chern number is an integer.

\section{Test for the method}

\subsection{Empty lattice}

In this subsection, we give appropriate settings for the Gaussian basis sets
to produce sufficiently
accurate results for empty lattices in 2D and 3D cases.
Further, using the
determined setting, we calculate band structures of typical
photonic crystals of both 2D and 3D as examples.  

Let us begin with the 2D empty lattice, in which the equation to be solved becomes a scalar
equation. The results for a square lattice are shown in Fig.~\ref{empty-lattice}(a) for two
different settings represented as solid and broken lines. 
For both of the settings, we use  $64$ basis elements on regularly aligned
grid points,
i.e., eight basis elements per
each direction. The broken line is obtained by setting the localization
factor in the $\mathrm{e}^{- r^2 / \alpha^2}$ to
 $\alpha = 0.87 d$ ($d$
being the spacing between two gird points) and the solid line is obtained
by setting it to $ \alpha = d$.
A good convergence for these settings requires
matrix elements up to the fifth-nearest-neighbor pairs of the
grid points. The necessary furthest pairs for matrix elements are
determined by the overlap length of the matrix $\Theta$ since the
convergence of the matrix elements of $\Theta$ is slower than those of $O$.

Within the shown frequency scale, the solid and broken lines overlap well
and the solid lines match perfectly with
the analytically derived dispersion, namely, $\omega = c k$ folded in the
first Brillouin zone.
However, at the $\Gamma$-point,
the lowest band does not converge to $\omega \rightarrow 0$, in
$k\rightarrow 0$ limit.
This mismatch comes from the fact that the
Gaussian basis elements should have sufficiently large overlaps between them in
order to compose a spatially homogeneous state, which is
expected to be realized in the $\omega\rightarrow 0$ and
$k\rightarrow 0$ limit.
On the other hand, the accuracy for higher frequency modes is improved by
increasing the number of basis
elements since higher modes vary more rapidly in space.
The failure in the $\omega\rightarrow{0}$ limit is usually
irrelevant in the discussion of topological states, 
since we usually focus on photonic gaps at finite $\omega$ to see topological
phenomena.

For the 3D case, 
the vector equations should be treated and the basis
set $\left\{ | \boldsymbol{\tilde{g}}_{ik}^{\lambda} \rangle \right\}$ is
used. Also in this case, 
the exact band
structure of the empty lattice is well restored by using $1,536$
basis elements,
i.e., 512  grid points with eight
 grid points per each direction, and three
 basis elements
$\left(  \lambda = 1, 2, 3\right)$ on each grid point. 
The results are shown in Fig.~\ref{empty-lattice}(b) for two
different settings. 
The broken line is obtained by setting the localization
factor in the $ | \boldsymbol{\tilde{g}}_{ik}^{\lambda} \rangle $[Eq.~\eqref{the-base}] to $\alpha =
 d$ and the solid
line is obtained by setting it to
$\alpha =  1.07 d$.
However, since the $|\boldsymbol{\tilde{g}}_{ik}^{\lambda} \rangle$ is the 
Gaussian function multiplied by the first-order polynomial, it extends
in a broader region than the genuine Gaussian function does.

\begin{figure}[htb]
\begin{center}
 \includegraphics[width=\columnwidth]{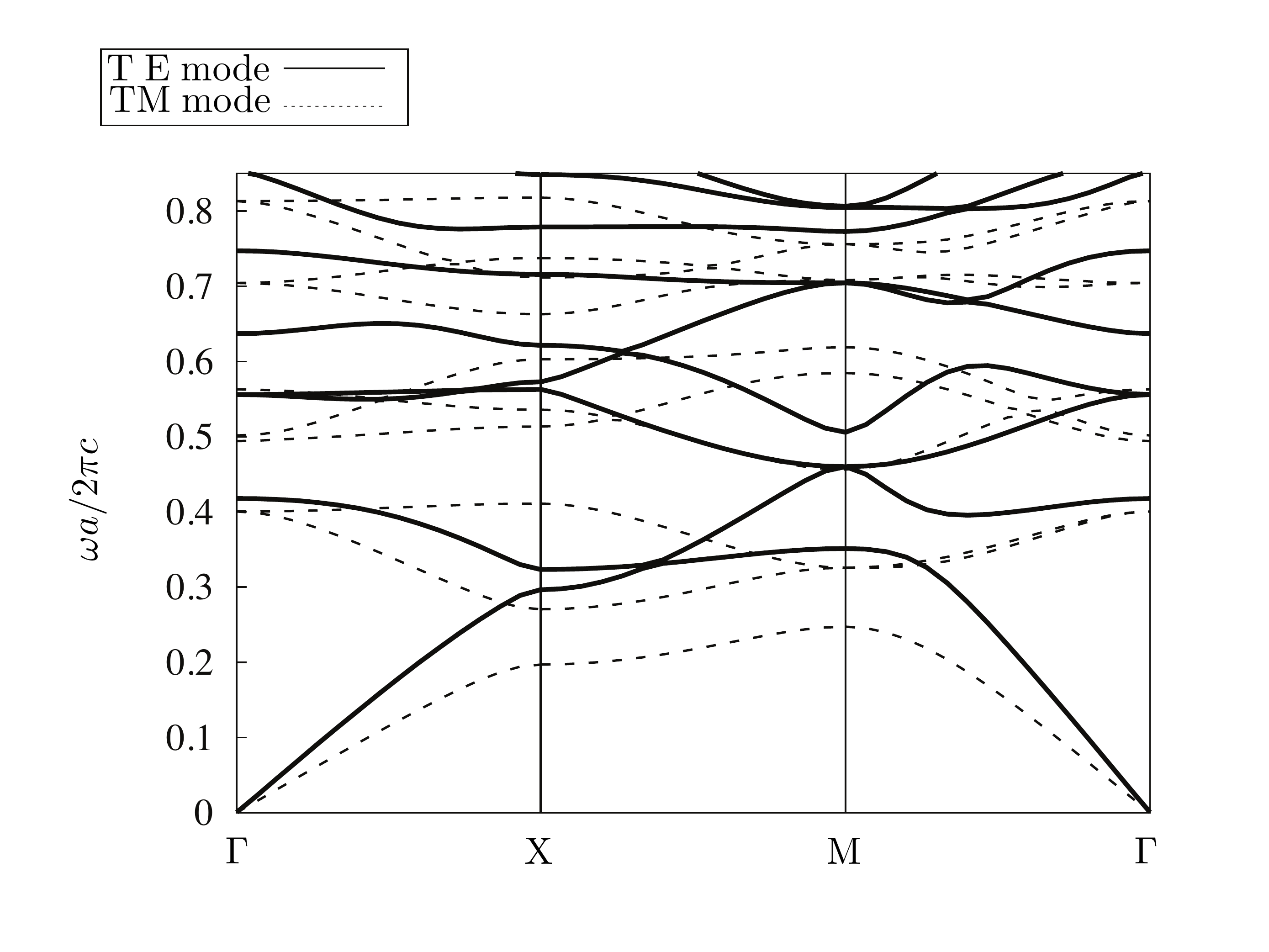}
 \caption{Calculated band structure of dielectric columns arranged in
 a square lattice in air with radius $r = 0.378 a$.
 We  set the relative permittivity $\varepsilon = 8.9$ in columns
 ({$\varepsilon = 1$} in air). Solid lines (dashed lines) represents TE (TM) mode
 dispersion. The calculation is performed using 64
  basis elements with
  $\alpha = d$. } \label{square-columns}
\end{center} 
\end{figure}

The difference of the first eigenvalues between $\alpha = d$ and 
$\alpha  = 1.07 d$
 near $\Gamma$-point is more prominent than the 2D case.
 We have confirmed that the 
 results are well converged and match perfectly to the dispersion,
 $\omega = c k$, folded in the first Brillouin zone in the shown frequency scale when we set $\alpha$ 
 more than
 $1.07 d$, which takes the matrix elements up to the $8$th
nearest neighbor points.
Yet the lowest eigenmode
 at $\Gamma$ point is never restored with the basis set
 $\{|\bm{\tilde{g}}_{ik}^{
 \lambda}\rangle\}$, due to the difficulty in
 representing a uniform field by
 $|\bm{\tilde{g}}_{ik}^{\lambda}\rangle$, since
 $|\bm{\tilde{g}}_{ij}^{ \lambda}\rangle$ is an odd function with respect to
 its  center. 

\subsection{The case of spatially modulated permittivity}

To begin with, we choose an array of dielectric columns arranged
on a square lattice as a sample system for the 2D
Gaussian basis set. We assume
the radius of the column to be $0.378 a$  with $a$ representing
the lattice spacing, and
the relative permittivity to be $\varepsilon = 8.9$ within the columns
($\varepsilon = 1.0$ out of the columns)\cite{Dobson1999}. The calculated dispersions for TE modes (solid
lines) and TM
modes (dashed lines) are shown in Fig.~\ref{square-columns}. The
calculation is done with the same setting as the empty lattice [the
solid line in Fig.~\ref{empty-lattice}(a)].
The numerical integration between Gaussian basis elements converges by
fining the integral mesh up to  $\alpha / 5$.
The precision of the results increases with the number of
 basis elements in one direction
$n_x$. In this system, the most of eigenvalues converge
in the shown scale before $n_x = 8$, with which the result in
Fig.~\ref{square-columns} is obtained. 

\begin{figure}[thbp]
\begin{center}
 \includegraphics[width=\columnwidth]{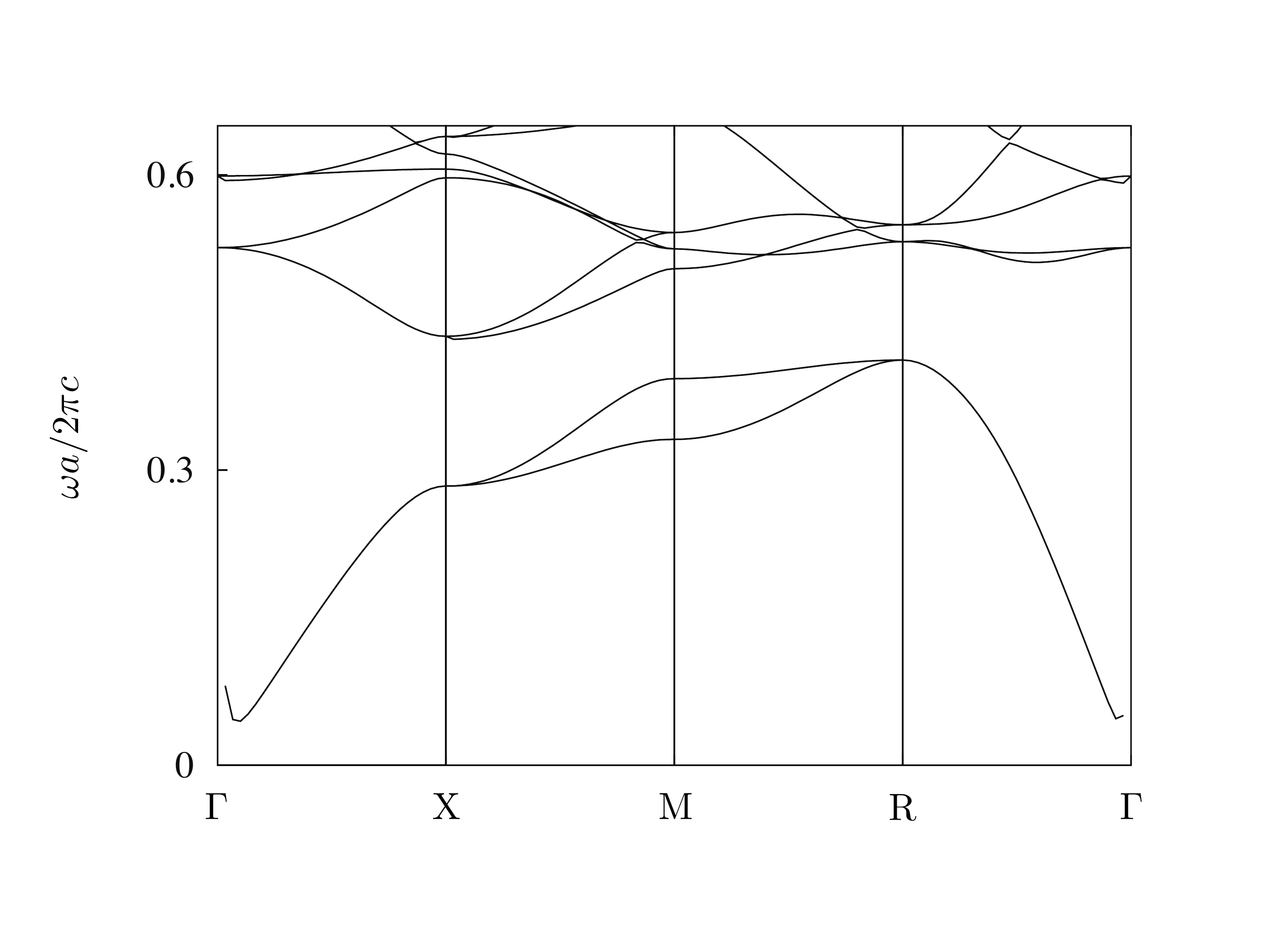}
 \caption{Calculated band structure of the
 three-dimensional square rod
 structure with the width of rod being $0.26a$ and the relative permittivity $\varepsilon$ =
 $13.0$. The calculation is performed using 1,536
 basis elements with $\alpha = 1.07 d$. } \label{cubic-scaffold}
\end{center} 
\end{figure}

In this case, the radius of the column $r = 0.378 a$ is relatively large
compared to
the size of the unit cell. Therefore the structure is smooth and well
restored with the integral mesh $\alpha / 5$ for $n_x = 8$. However,
when the radius of the column is smaller, the basis set
has to detect a finer
structure in integration and we need a larger number of
basis elements, which implies slower
convergence.  Another potential source of bad convergence is a
sharp structure such as a corner of a rectangular rod, and we should be
careful when such a structure is treated in the present method.

Next, we choose a simple scaffold structure, namely a
three-dimensional square rod structure studied by S\"oz\"uer and
Haus\cite{Sozuer1993}, as an example system for the 3D Gaussian basis set. We assume the width of the rod as $0.26 a $ and
the relative permittivity $\varepsilon = 13.0$, which is the same situation as in
the study of Dobson \textit{et al.} \cite{Dobson2000}. The result shown in Fig.~\ref{cubic-scaffold} is obtained with the same
setting as in the empty lattice case [the
solid line in Fig.~\ref{empty-lattice}(b)].
As known, there exists full gap between the
 fifth and  sixth bands.
As in the case of the empty lattice, there is
difficulty in the $\omega\rightarrow 0$ and $k\rightarrow 0$ limit,
however, except that point, the global profile of the calculated band
structure, such as the existence of the full gap, is consistent with
that of the previous works.

\section{Section Chern numbers and edge states in a 3D photonic crystal}
In this section, we demonstrate the topologically
protected edge states of EM waves associated with the
finite section Chern number in a 3D photonic crystal with broken SIS. 

In a 3D periodic system, the wave vector has three
components, $k_1$,
$k_2$, and $k_3$, and the section Chern number is defined using two of
them, regarding the remaining one as a free parameter.
If we fix $k_3$, the section
Chern number $C_{n} (k_3)$ is defined as 
\begin{equation}
\begin{aligned}
 C_{n} (k_3) &= \frac{1}{2\pi \mathrm{i}} \int \mathrm{d}k_1 \mathrm{d}k_2 \thickspace B_{n,12}
 (\bm{k}), \\
 B_{n,12} (\bm{k}) &= \partial_{k_1} \mathcal{A}_{n,2} \left( \bm{k} \right) -
  \partial_{k_2} \mathcal{A}_{n,1} \left( \bm{k} \right),
  \label{secChern_def} \\
%
 \mathcal{A}_{n,i} (\bm{k}) &= \langle \psi_n (\bm{k}) |
 \partial_{{k}_i} | \psi_n (\bm{k}) \rangle, 
\end{aligned}
\end{equation} 
and those defined by fixing $k_1$ or $k_2$ are similarly given by
permutations of the components
$(k_1,k_2,k_3)\rightarrow (k_2,k_3,k_1) \rightarrow (k_3,k_1,k_2)$. In order to make $C_{n} (k_3)$
well-defined, the gap should remain finite over the
entire $k_3$ constant plane. In general, the section Chern
number takes any value, but the symmetry of a given system induces
restriction. If a system has TRS, the eigenstate for $\bm{k}$ is
 related to that for $-\bm{k}$ by the bosonic time-reversal
 operation $\mathcal{T}$, as
 $\mathcal{T}|\psi_n(\bm{k})\rangle = |\psi_n(-\bm{k})\rangle^*$, which
 leads to $\bm{B}_n (-\bm{k})=-\bm{B}_n (\bm{k})$. Then,
 we have $C_{n} (k_3)=0$ for $k_3=0$ and $\pi$, because of
 $B_n (-k_1,-k_2,0)=-B_n (k_1,k_2,0)$ and
 $B_n (-k_1,-k_2,-\pi(=\negthickspace \pi))=-B_n (k_1,k_2,\pi)$, and the
 definition Eq.~\eqref{secChern_def}. If there is SIS in
 addition to TRS, an additional restriction may  arise, that is, SIS gives
$|\psi_n(-\bm{k})\rangle=|\psi_n(\bm{k})\rangle$, which indicates 
$\bm{B}_n (-\bm{k})=\bm{B}_n (\bm{k})$, and if this relation is
 combined with $\bm{B}_n (-\bm{k})=-\bm{B}_n (\bm{k})$ required
 by TRS, we finally obtain
 $\bm{B}_n (\bm{k})=-\bm{B}_n (\bm{k})=0$. 
 Therefore, SIS has to be broken for a nonzero section Chern number if the
 system has TRS.

Owing to the topological nature of $C_{n} (k_3)$, when $C_{n} (k_3)$ changes
as a function of $k_3$, there should be a gap closing point somewhere
on the corresponding $k_3$ constant plane. On the other hand, as we
have noted, the section Chern number is always zero at $k_3=0$ and
$\pi$ with TRS. These facts indicate that the finite section Chern
number requires existence of a gap closing point. Typically, gap
closing occurs on isolated points in the Brillouin zone and the linear
dispersion appears around those degeneracy points. A degeneracy point
with linear dispersion is named as a Weyl point. 
Note that $k_z$ appearing in the
section Chern number $C_{n} \left(k_z\right)$
is related to the fixed $k_z$ of an incident wave on the photonic crystal.

\subsection{Honeycomb array of air hole columns} 

\begin{figure}[htb]
\begin{center}
 \includegraphics[width=\columnwidth]{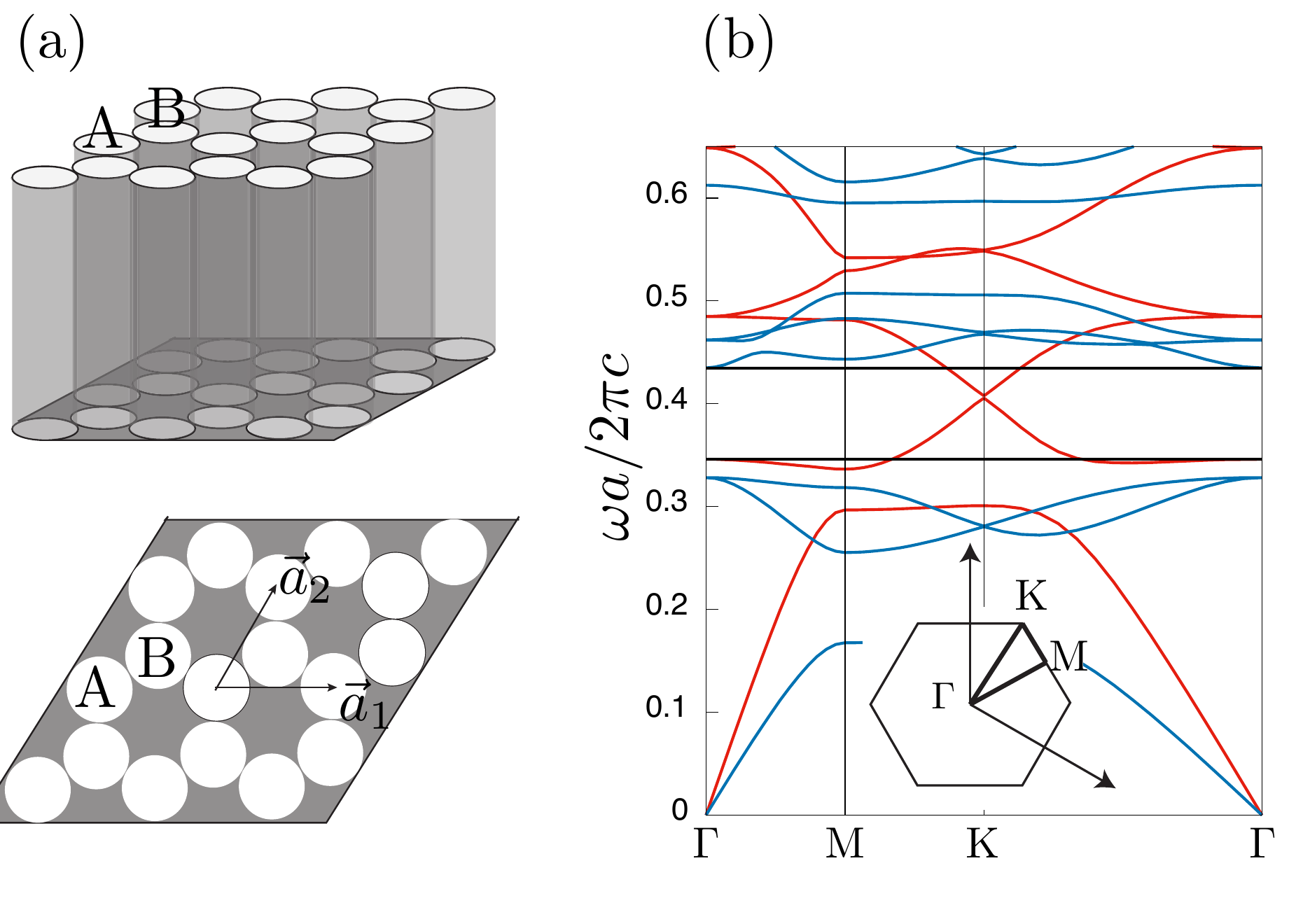}
 \caption{(Color online) (a) Honeycomb array of hollow columns in
 a dielectric
 material. (b) Two-dimensional dispersion
 of the photonic crystal. We use permittivity $\varepsilon = 20.0$, the radius
 of hollow column $r = \sqrt{3}a / 6$. The band structure of TE
 modes (red line) has a distinctive
 degeneracy point, which is clear (indicated by black lines) in the $k_z=0$
 plane among both of TE and TM modes.}
 \label{hollow-columns}  
\end{center} 
\end{figure} 

Now, we consider to realize Weyl points in photonic
crystals. In specific, we
need a system in which the Weyl point related
physics is readily accessible. 
More specifically, in order
to observe chiral edge modes associated with the finite section Chern
number $C_{n}(k_3)$, it is required to have a clear gap on the entire
two-dimensional plane with fixed $k_3$. Since a clear gap is required
only on a fixed $k_3$ plane, a possible strategy to achieve our goal is
to start with a
two-dimensional system having ``pseudo-gap,'' a
frequency region filled with only a few bands, and then to apply appropriate
three-dimensional modifications to the system. Because we are handling
photonic crystals,
the band structure can be modified by controlling the background
medium. However, our experience tells us that careless modifications
prone to fail, namely, even if Weyl points are successively generated,
they are often masked by the other dispersive bands. Hence, the careful
design is important and this is what we discuss in the following. 

In order to realize easily accessible Weyl points, we consider a
photonic crystal of hollow columns aligned on the honeycomb lattice as
depicted in Fig.~\ref{hollow-columns}. The band structure of this system
has some similarity to that of the triangular lattice consisting of
dielectric columns\cite{Plihal1991}. This is because the complementary region of the
hollow honeycomb lattice composes a triangular lattice. Despite this
similarity, we choose to use the hollow honeycomb lattice since it turns
out that the existence of the sublattice structure of the honeycomb
lattice is advantageous in the following discussion. Without
three-dimensional modulation, the second and third TE modes of the
hollow honeycomb lattice form 2D Dirac cones at K and K' points, and TM
modes do not mask these TE Dirac cones.

Owing to SIS of the hollow honeycomb lattice, these TE Dirac cones are
actually line degeneracies extending in the $k_z$ direction in the
three-dimensional Brillouin zone. Then, SIS should be broken to
transform the line degeneracy into Weyl points, and therefore, we reshape
each hollow column into a hollow helix. Note that it is possible to
modulate columns differently on sublattices A and B of the honeycomb lattice.

\subsection{Weyl points and change in section Chern number}
\begin{figure}[tbp]
\begin{center}
 \includegraphics[width=\columnwidth]{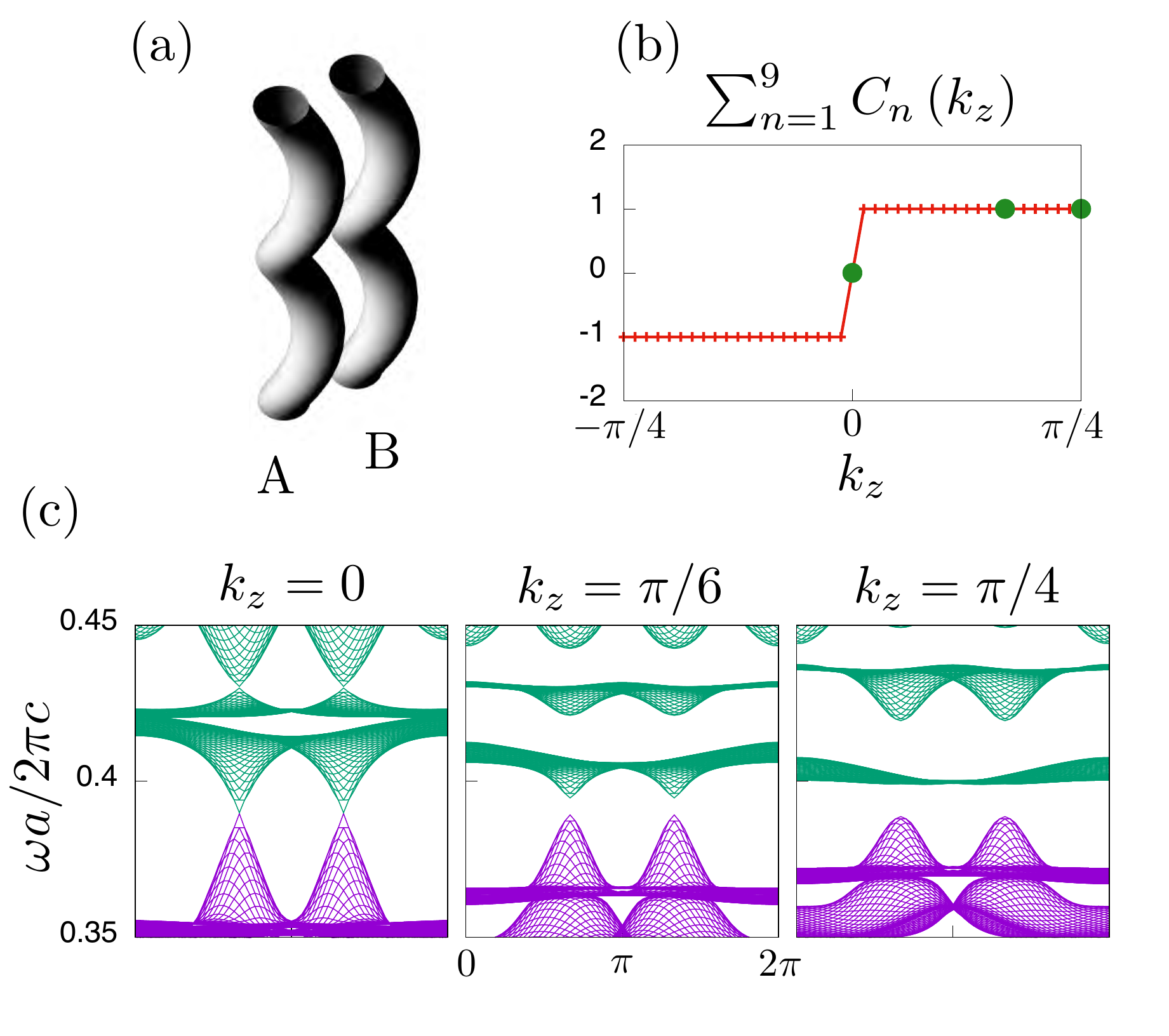}
  \caption{(Color online) (a) Columns are modulated into helices with the same twists
 for sublattices A and B. The radius of the twisting circle is taken to be
 $r_h = 0.05 a$ and it circles one time with translation in
 the $z$ direction by a unit length $a$. The system does not have SIS but is
 invariant under the twofold
 rotation around the $z$ axis and the
 axis perpendicular to the $z$ axis. (b) The
 value of the total section Chern number summed up to the 9th band
 for each $k_z$ $\left( - \pi/4 \le k_z \le \pi/4 \right)$. (c) Bulk
 band dispersions for several values of $k_z$. The 1-9th bands are represented
 by purple lines.}
 \label{bulk-equal}  
\end{center} 
\end{figure}

Now, we consider to apply the same twist on the both hollows at
sublattices A and B (Fig.~\ref{bulk-equal}). By transforming columns into
helices, the SIS is broken and the line degeneracy is lifted except K
and K' points, which implies emergence of the Weyl points. This 
emergence is explained in terms of the point group symmetry as
follows. Before the twist is applied, the system has $D_{6h}$ symmetry,
and the group of $k$ is $C_{3v}$ on the line parallel to
the $k_z$ axis passing through K and K' points, whereas
it is $D_{3h}$ for K and K' points. Since both of $C_{3v}$ and $D_{3h}$
contain two-dimensional representations, twofold degeneracy is allowed on the
whole line, which
supports existence of the line degeneracy without twist. Then, after
applying the twist, the reflection symmetry with respect to a plane
including the $z$ axis is broken and the group of $k$ on the line parallel to
the $k_z$ axis passing through K and K' points turns to $C_3$, except K and
K' points. This implies the line degeneracy dissolves except K and K'
points since $C_3$ has no multidimensional representation. For K and K'
points, the group of $k$ changes from $D_{3h}$ to $D_3$. By the
compatibility relation, 2D representations of $D_{3h}$ are connected to
2D representations of $D_3$, and therefore, the degeneracies at K and K'
points survive even with the twist.

The surviving degeneracy points are
two Weyl points with the same chirality since these points are related
by two-fold rotation around the $z$ axis. Then, the section Chern number
$C_{n}(k_z)$ changes by $\pm 2$ when $k_z$ crosses $k_z=0$, because of \textit{two}
Weyl points on the $k_z=0$ plane [Figs.~\ref{bulk-equal}(b) and
\ref{bulk-equal}(c)]. The exactly
same argument also applies for the $k_z=\pi$ plane, but it is not possible
to resolve Weyl points on the $k_z=\pi$ plane, since the bands on which we
focus merge into the other bands as we increase $k_z$ from $0$ to
$\pi$. Note that the observation of $\pm 2$ jump in $C_{n}(k_z)$ at $k_z=0$
does not contradict with the former statement that $C_{n}(0)$ should be
$0$, since at that time we had assumed that the gap is finite on entire
$k_z=0$ plane, and this assumption is invalid in this case.

\begin{figure}[tbp]
\begin{center}
 \includegraphics[width=\columnwidth]{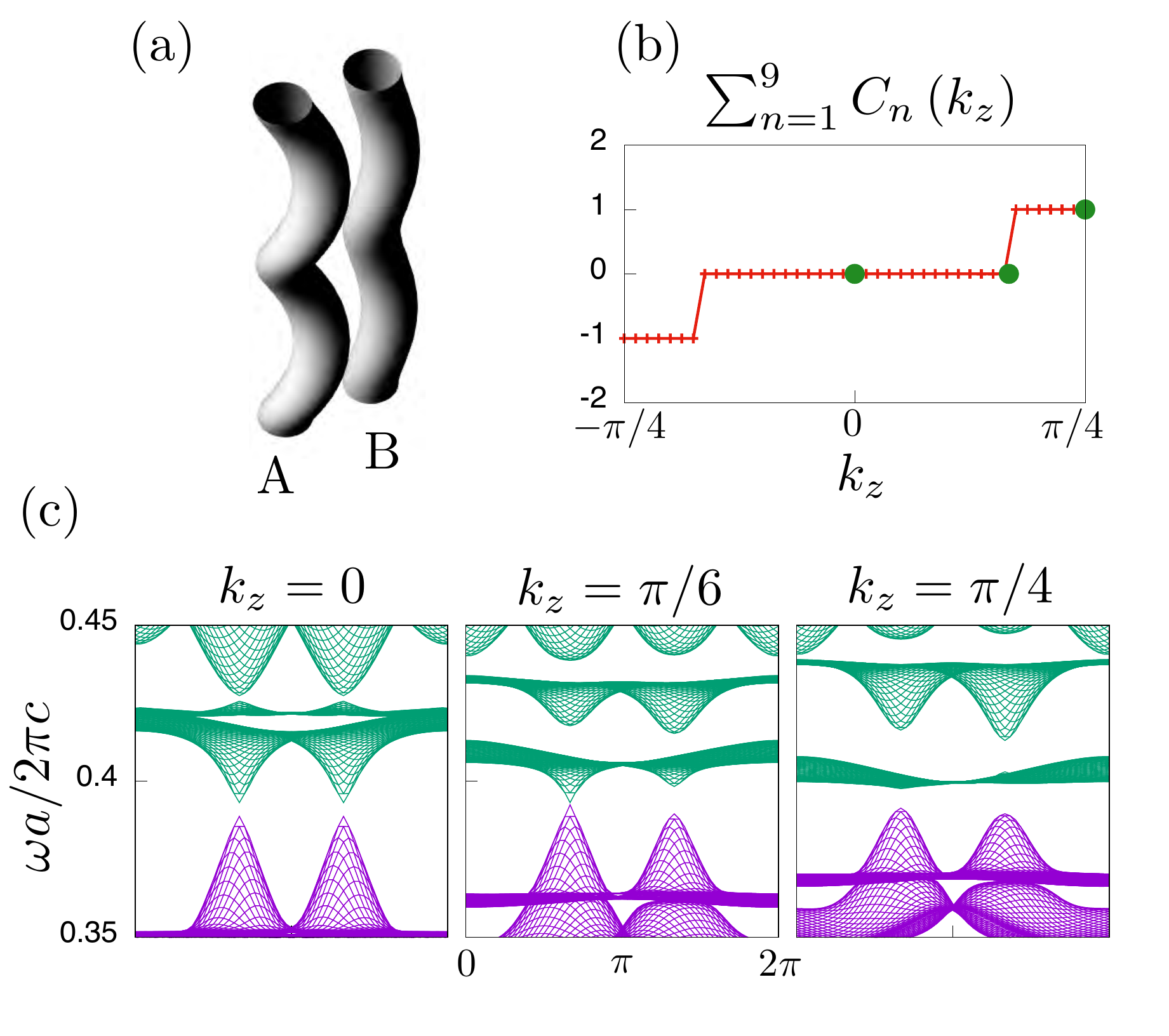}
  \caption{(Color online) (a) Different twists for sublattices A and B, with the twisting
 circle for helix A  being $r_{h,A} = 0.05 a$
 and for helix B  being $r_{h,B} = 0.03 a$. (b) The
 value of the total section Chern number summed up to the ninth band
 for each $k_z$ $\left( -\pi/4 \le k_z \le \pi/4 \right)$. (c) Bulk
 band dispersions for several values of $k_z$. The first to ninth bands are represented
 by purple lines.}\label{bulk-differentr} 
\end{center} 
\end{figure}

Next, we consider to change the twist of helix only for the hollow on
the sublattice B [Fig.~\ref{bulk-differentr}(a)]. With this modification,
the twofold rotational symmetry around the $k_z$ axis and
the axis perpendicular to $k_z$ axis are broken and then,
there is no reason to have degeneracy on the $k_z=0$ plane. However, the
Weyl points found in Fig.~\ref{bulk-equal}(c) do not simply disappear
because Weyl points are topologically stable and only disappear by
annihilating as a pair of Weyl points with opposite
chiralities
\cite{Nielsen81}. Our numerical calculation shows that the Weyl points move to $\pm k_z$
directions as we increase the difference of the radius of the
helices. When one of the Weyl points moves in $+k_z$ direction, the
other one moves in $-k_z$ direction due to TRS. In this case, the
dispersion is fully gapped on the $k_z=0$ plane, and we have $C_{n}(0)=0$,
which is consistent with the former statement. The section Chern number
becomes finite for large enough $k_z$ as expected from the existence of
the Weyl points. 

\subsection{Edge modes}
\begin{figure}[tbp]
\begin{center}
 \includegraphics[width=\columnwidth]{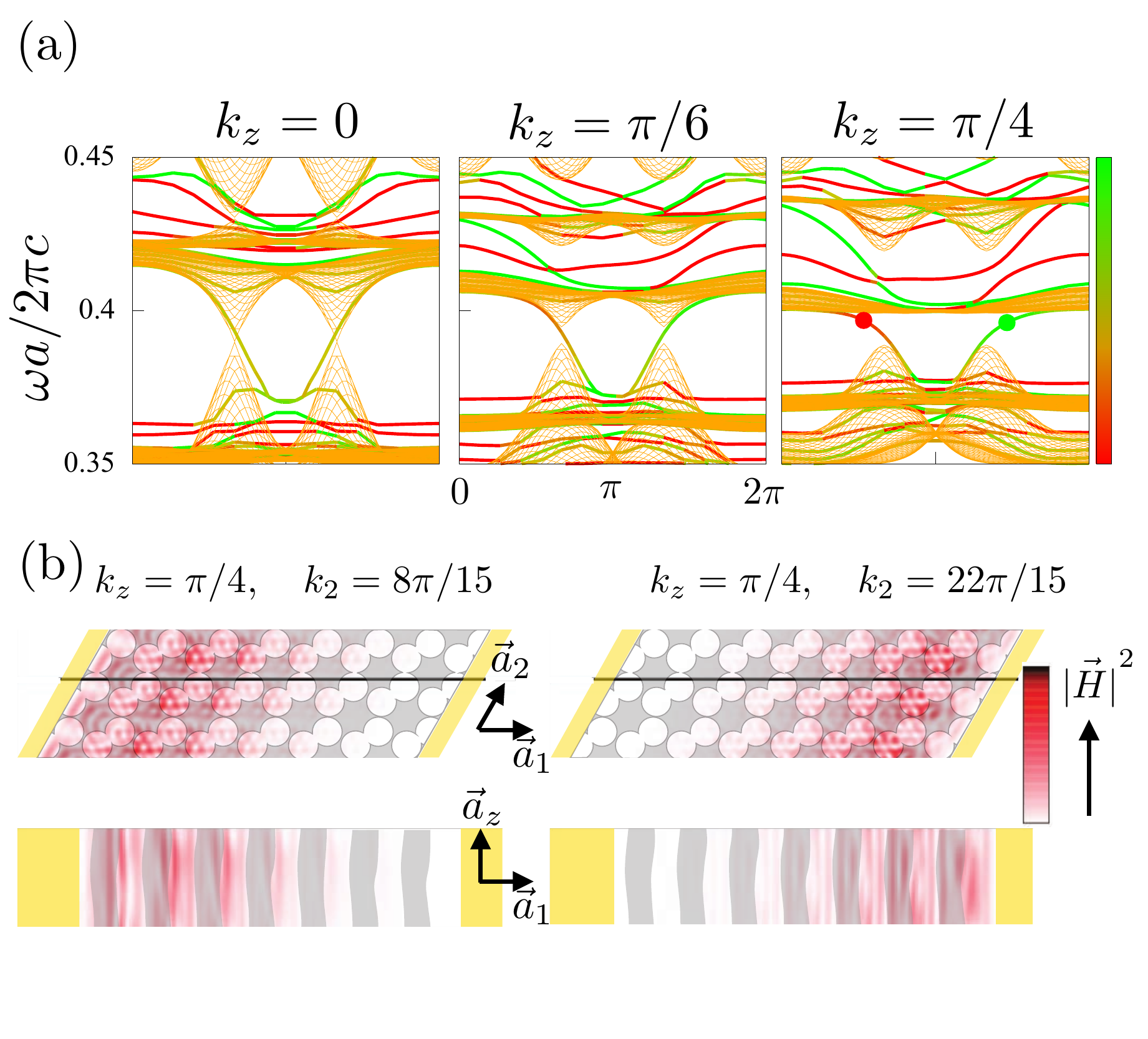}
 \caption{(Color online) (a) The dispersion curve (color spectrum) for a finite width
 system (eight unit cells) that corresponds to Fig.~\ref{bulk-equal} in bulk,
 plotted with projected bulk dispersion (yellow lines). The
 color spectrum indicates on which side the eigenstate 
 inclines by red (left) - yellow (middle) - green (right). 
(b) The profile  of the $n$th
 eigenstate for $k_z = \pi/4$, $k_2 = (4/15)
 2 \pi$ and $k_2 = (11/15) 2 \pi$, which are indicated by
 the points in the right most panel of (a).}
 \label{edge-mode-same} 
\end{center} 
\end{figure} 

In order to investigate edge modes, we perform calculations using the
developed Gaussian basis elements on 
the photonic crystal truncated in the $\vec{a}_1$ direction [see
Fig.~\ref{edge-mode-same}(b) for the definition of $\vec{a}_1$] with
finite width (eight unit cells in specific). At the interface, we place a
material with a smaller dielectric constant so as to prevent the light from
evading to outside. This is necessary because the momentum/frequency
region we are focusing on lies above the light cone (for the same
dielectric constant with the hollow region), which means that
the light is able to escape from the system.

\begin{figure}[tbp]
\begin{center}
 \includegraphics[width=\columnwidth]{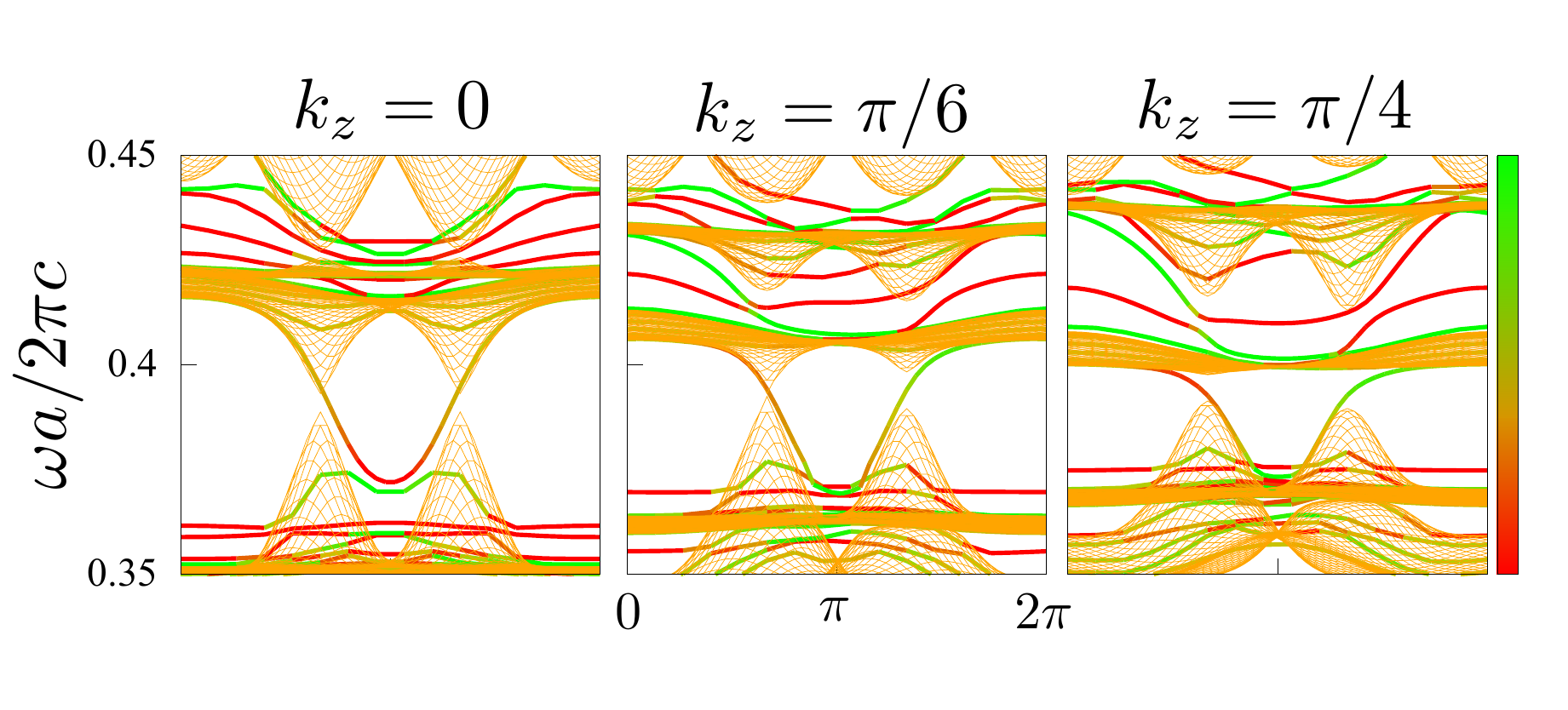}
 \caption{(Color online) The dispersion curve (with color spectrum) for a finite width
 system (eight unit cells) that corresponds to
 Fig.~\ref{bulk-differentr} in bulk, plotted with projected bulk
 dispersion (yellow lines). See the caption of
 Fig.~\ref{edge-mode-same} for the description of the color spectrum.}
 \label{edge-mode-differentr}
\end{center} 
\end{figure}

In Figs.~\ref{edge-mode-same} and
\ref{edge-mode-differentr}, the dispersion relations of the finite
width system are compared with the bulk dispersions projected on the
surface for several values of $k_z$. We find several bands apart from
the bulk contribution, which signals the existence of edge modes. In
fact, it is confirmed that the non-bulk bands are localized at the
interfaces by examining the eigenvector for each mode. The eigenvectors
also tell us on which side, left or right, each mode localizes. In
Figs.~\ref{edge-mode-same} and \ref{edge-mode-differentr}, the red color corresponds to the states
on the left edge, while the green color to the states on the right
edge. The color spectrum is determined by the moment of the eigenfield
$\langle r_1 \rangle_{\scriptscriptstyle \boldsymbol{H}_{n,k}} = \int
\mathrm{d}^3 r  \thickspace r_1 \vec{H}_{n,k}^2 \negthinspace
\left(\boldsymbol{r}\right)$. 
As we change $k_z$, the way that the edge modes connect bulk bands
changes in accordance with the change of the sum $\sum_{n=1}^9 C_{n}(k_z)$, which confirms the
bulk-edge correspondence. The same argument also applies to gaps other
than the 9th gap with nonzero total section Chern number.
When the fixed momentum is conserved for $k_z$
with
$\sum_{n=1}^{i} C_{n} \negthinspace (k_z)$ being nonzero,
the wave propagates unidirectionally along the interface.

This photonic crystal has the structure
similar to the honeycomb array of twisted waveguides studied by
Rechtsman \textit{et al.}\cite{Rechtsman2013}, where evanescently
coupled modes are considered. In the system of evanescently coupled
waveguides, the $z$ axis, which is the propagation direction of a waveguide mode, is
regarded as the \textit{temporal} axis
and the system is spatially two-dimensional. There, the twist of a waveguide
becomes a temporal periodic modulation for waveguide modes and 
the appearance of the chiral edge modes is attributed to
the Chern number for the \textit{Floquet} band structure.
On the other hand, we consider, in this paper,
TE-like modes in the spatially three-dimensional photonic crystal and
relate the unidirectional propagation of them to the finite value of
section Chern numbers.

Depending on the dielectric constant of the material, there might exist some wave guide
modes near the frequency of the Weyl point and it leads to coupling
between edge modes and wave guide modes that extend into bulk. In such a case, it is possible to suppress the mixing
between the edge modes and the wave guide modes by shifting the relative
positions of them in the frequency space. The relative shift can be
achieved by changing the pitch of the helix while keeping its twist
unchanged because the dispersion of wave guide modes is easily affected
by the periodic length in the $z$ direction, whereas that of TE-like modes is not affected so much.

\section{Summary}

We proposed the Gaussian basis sets for the calculation of EM fields in
2D and 3D photonic crystals.
In the formulation with spatially localized basis elements, the wave number appears as a boundary
condition in the eigenequation and 
the eigenequation strictly becomes periodic
in the wave number, which is advantageous for the Chern number calculation.
In addition, the localized property of the Gaussian
basis element becomes
effective for the consideration of finite size or interface effects,
which usually requires larger systems, 
by utilizing iterative algorithms.
Besides, in three-dimensional cases, the Gaussian
basis element can easily
be accommodated
to the divergence-free constraint of the Maxwell equations due to its
simplicity for differentiation operation.
For further improvement, it would be expected to optimize the mesh
alignment or the localization factor of the Gaussian
basis elements for each structure.  

We demonstrated the bulk-edge relation between the section Chern number
and chiral edge modes in the SIS broken 3D photonic crystal with
TRS. It was confirmed that chiral edge modes in the
$n$th gap at each $k_z$ reflects the
total of the section Chern number
below the $n$th gap $C_{n} \left(k_z\right)$.
A system with a finite section Chern number is
expected to lead to some applications for wave-packet dynamics. For
simplicity, we take a situation that the Weyl points exist on the
$k_z=0$ plane. 
When a wave packet is injected into the finite width system of the
SIS broken 3D photonic crystal, a wave packet composed of the Bloch states with
positive $k_z$ does propagate on the one
side, but does not on the other side. If we use the Bloch states with
negative $k_z$, the side that allows propagation changes. In this way,
the system is useful to filter wave packets with fixed sign of $k_z$. 

\begin{acknowledgments}
We thank S.~Takahashi and S.~Iwamoto for fruitful discussions.
This work is partly supported by Grants-in-Aid for Scientific Research,
Nos. 26247064, 25107005, and 16K13845 from JSPS. 
The computation in this work has been done using the facilities of the
Supercomputer Center, the institute for Solid State Physics, the
University of Tokyo.
\end{acknowledgments}

\bibliographystyle{apsrev4-1}
\bibliography{gaussian}
\end{document}